\pdfoutput=1
\documentclass{JINST}

\title{Detection of low energy antimatter with emulsions}

\author{
S. Aghion$^{a,b}$,
A. Ariga$^c$,
T. Ariga$^c$,
M. Bollani$^{d}$,
E. Dei Cas$^{b,e}$,
A. Ereditato$^c$,
C. Evans$^{a,b}$, 
R. Ferragut$^{a,b}$,
M. Giammarchi$^b$,
C. Pistillo$^c$\thanks{Corresponding author.}~, 
M. Rom\'{e}$^{b,e}$,
S. Sala$^{b,e}$,
P. Scampoli$^{c,f}$

\\
\llap{$^a$}LNESS Laboratory and Dipartimento di Fisica, Politecnico di Milano, Via Anzani 42, 22100 Como, Italy\\
\llap{$^b$}Istituto Nazionale di Fisica Nucleare, Sez. di Milano, Via Celoria 16, 20133 Milano, Italy\\
\llap{$^c$}Albert Einstein Center for Fundamental Physics, Laboratory for High Energy Physics, University of Bern, Sidlerstrasse 5, 3012 Bern, Switzerland\\
\llap{$^d$}Istituto di Fotonica e Nanotecnologie del CNR, LNESS Laboratory, via Anzani 42, 22100 Como, Italy\\
\llap{$^e$}Dipartimento di Fisica, Universit\`{a} degli Studi di Milano, via Celoria 16, 20133 Milano, Italy\\ 
\llap{$^f$}Dipartimento di Fisica\lq{\lq{Ettore Pancini}\rq}\rq, Universit\`{a} di Napoli Federico II, Complesso Universitario di Monte S. Angelo, 80126 Napoli, Italy\\

E-mail: \email{ciro.pistillo@lhep.unibe.ch}}

\abstract{Emulsion detectors feature a very high position resolution and consequently represent an ideal device when particle detection is required at the micrometric scale. This is the case of quantum interferometry studies with antimatter, where micrometric fringes have to be measured. In this framework, we designed and realized a new emulsion based detector characterized by a gel enriched in terms of silver bromide crystal contents poured on a glass plate. We tested the sensitivity of such a detector to low energy positrons in the range 10-20 keV.
The obtained results prove that nuclear emulsions are highly efficient at detecting positrons at these energies. This achievement paves the way to perform matter-wave interferometry with positrons using this technology.}


\begin{document}

\section{Introduction}

The use of nuclear emulsions as tracking detector in particle physics dates back to more than a century ago.
From the nineties onwards, a renewed interest was triggered by neutrino physics, thanks to parallel technological advances in the field of emulsion production and automatic scanning ~\cite{emul_review2}. More recently, their exploitation in additional fields of research has begun.
Among those, an emulsion based detector is particularly a promising option to measure the free fall of antihydrogen atoms in the Earth's gravitational field at CERN~\cite{AEgIS_emulsion1}. Its capability to reconstruct antiproton annihilations with micrometric resolution has also been demonstrated~\cite{AEgIS_emulsion2}.
Furthermore, a proof of the detection principle (the observation of the shadow pattern produced by antiprotons propagating through a moir\'e  deflectometer coupled to an emulsion detector) was successfully conducted~\cite{AEgIS_minimoire}.\\
\indent 
In this paper we discuss a further application of such a device for antimatter wave detection in the wake of a recent paper where Sala et al.~\cite{Sala_antimatter} propose the Talbot-Lau set up to carry out matter-wave interferometry as the best configuration for decaying particles, also in case of a Gaussian distribution of the particle energy, which describes quite realistically the actual positron and antiproton beams. 
Emulsion detectors represent the optimal detector for this purpose, being capable of resolving interferometric patterns at the micrometric scale.
Our work proves that nuclear emulsions feature high detecting efficiency for low energy antiparticles, namely positrons in the range  10-20 keV. \\
\indent
The  experimental methodology described in this paper paves the way for matter-wave interferometry with positrons and, more generally, can be applied for the study of fundamental symmetries and interactions, such as CPT and the Weak Equivalence Principle, towards a future scientific initiative (QUPLAS: QUantum interferometry and gravitation with Positrons and LASers). The main working principle of QUPLAS involves using the positron beam described here, micrometric gratings and the emulsion detector in order to study quantum interferometry of positrons as a straightforward application of this technique. In a subsequent phase, the insertion of transmission targets in the positron beam, and of a suitable excitation laser will make possible to study positronium (Ps) quantum interferometry and the Ps fall in the Earth gravitational field.

\section{Materials and Methods}

\begin{figure}[h]
\begin{center}
\includegraphics[width=0.60\textwidth]{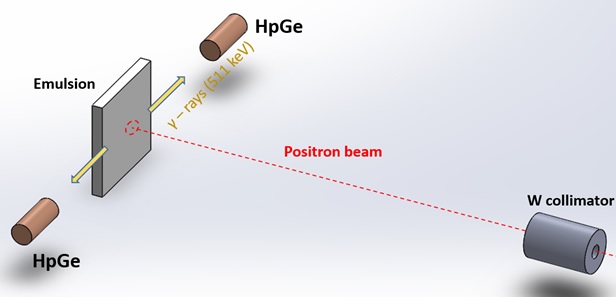}
\caption{A sketch of the experimental setup}
\label{drawing}
\end{center}
\end{figure}

Nuclear emulsions are composed by silver-bromide microcrystals with diameter of $\sim0.2$ $\mu$m embedded in a gelatine matrix.  
The passage of  ionizing radiation across a crystal produces a latent image which, after a chemical development process, leads to the creation of  a $\sim1~\mu$m silver grain visible by means of an optical microscope. A review on the technology can be found in \cite{emul_review2}. \\
\indent
The emulsion gel for low energy positron detection was produced at the Nagoya University in Japan. It features an enriched content of silver bromide crystals ($\sim$55\% in volume) and low background\footnote{Referring to the emulsions produced by the Fuji company in Japan for the OPERA neutrino experiment with a silver bromide content of $\sim$30\% and a background of  $\sim$3 grains/1000 $\mu$m$^{3}$ .}, defined in terms of the number of thermal induced grains, that is $\sim$1-2 grains/1000 $\mu$m$^{3}$.
Emulsion detectors are then produced by pouring the gel on a glass plate at the Laboratory of High Energy Physics of the University of Bern. The glass support is needed for its small thermal expansion coefficient, about one order of magnitude lower than the usual emulsion plastic supports. In fact, this is a relevant point in view of the detection of micrometric fringes over surfaces of several mm$^2$. Besides the local sub-micrometric precision provided by the emulsion detectors, it is crucial to keep an overall stability at the micrometric level after the chemical development of the emulsions, which could introduce significant deformations in the reconstructed pattern.
A 1.0$\pm$0.1 $\mu$m gelatine layer on the emulsion surface protects the sensitive layer against minor damages. It also prevents background enhancement during handling and transportation operations. Furthermore, glycerine is added to the gel at a concentration of 1.5\% to allow operation in vacuum, as described in~\cite{AEgIS_emulsion1}. The densities of the emulsion layer and the protective  layer were $4.0\pm0.3$ $g/$cm$^{3}$ and $1.3 \pm0.1$ $g/$cm$^{3}$ respectively.\\
\indent
The sensitivity of the emulsions to low energy particles was studied employing mono-energetic positron beams at the L-NESS laboratory in Como (Italy). This facility allows the tuning of the positron implantation energy from 0.1 to 20 keV. The positrons are emitted by a $^{22}$Na source and moderated by a monocrystalline tungsten foil. The beam operates in vacuum at pressures between 10$^{-8}$-10$^{-6}$  mbar. During the emulsion exposure, the gamma radiation carrying the information of the positron-electron annihilation process was simultaneously collected by two high-purity germanium detectors (HPGe) both located at a distance of 9 cm from the target (see  Figure~\ref{drawing}).\\

\indent
The positron rate at the target during the measurement at 18 keV was $(7.6\pm0.9)\times 10^{3}$~s$^{-1}$.  This estimate was performed by measuring the intrinsic efficiency of the HPGe detectors for photons of 511 keV (27$\pm1\%$) using a calibrated $^{22}$Na source and taking into account the solid
angle of the detector, the attenuation of the vacuum chamber, the emulsion and the aluminum holder, the positron backscattering, etc. After the exposures to positrons and the chemical development, the emulsions were analysed at the automatic microscopes of the scanning laboratory of the University of Bern, where fast data reconstruction based on GPU technology of large area emulsion surfaces is possible thanks to a recent development~\cite{ATmic}. The reconstruction software was optimized for the present study, where positrons with energies at the keV scale have to be detected, to maximize the performances in terms of surface detection and cluster identification. Depending on the actual particle energy, the range in emulsion is limited to a few microns or even less than 1 $\mu$m, so that only 1-2 crystals are involved. Consequently it is not possible to reconstruct tridimensional tracks, as for the standard readout of emulsions exposed to more energetic particles, and isolated clusters need to be identified.

\section{Exposure and Results}
\begin{figure}[h]
\begin{center}
\includegraphics[width=0.7\textwidth]{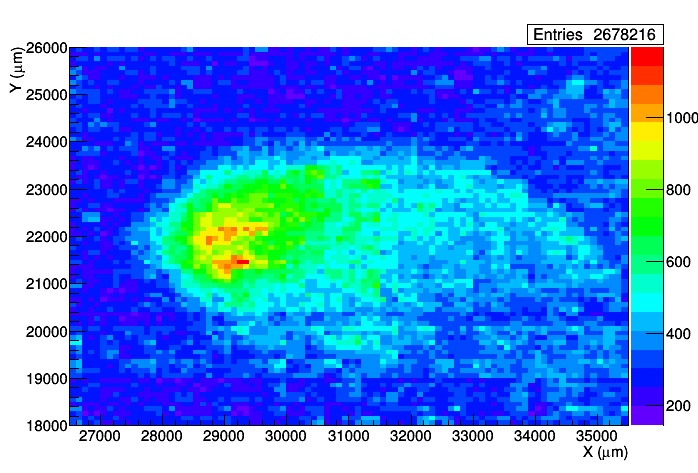}
\caption{Beam profile for 15 keV positrons. Colors are relative to the number of positrons in $10^{4} \mu$m$^{2}$ emulsion surface}
\label{15keV_2D}
\end{center}
\end{figure}

Emulsion exposures were performed at six different positron implantation energies (3, 6, 9, 12, 15 and 18 keV). The exposure time was set to ($120\pm1$)s for each energy, in order to have a sufficiently high signal to background ratio. 
A nearly gaussian positron spot (FWHM about 2.4 mm) was delivered to the emulsions for each energy. Then an area of about $1.5\times 1.5$ cm$^{2}$ around each spot was fully scanned and processed. The positron beam profile at the energy of 15 keV is shown in Figure~\ref{15keV_2D} where different colours account for the number of positrons in $10^{4} \mu$m$^{2}$ emulsion surface.\\

\begin{figure}[t]
\begin{center}
\includegraphics[width=0.7\textwidth]{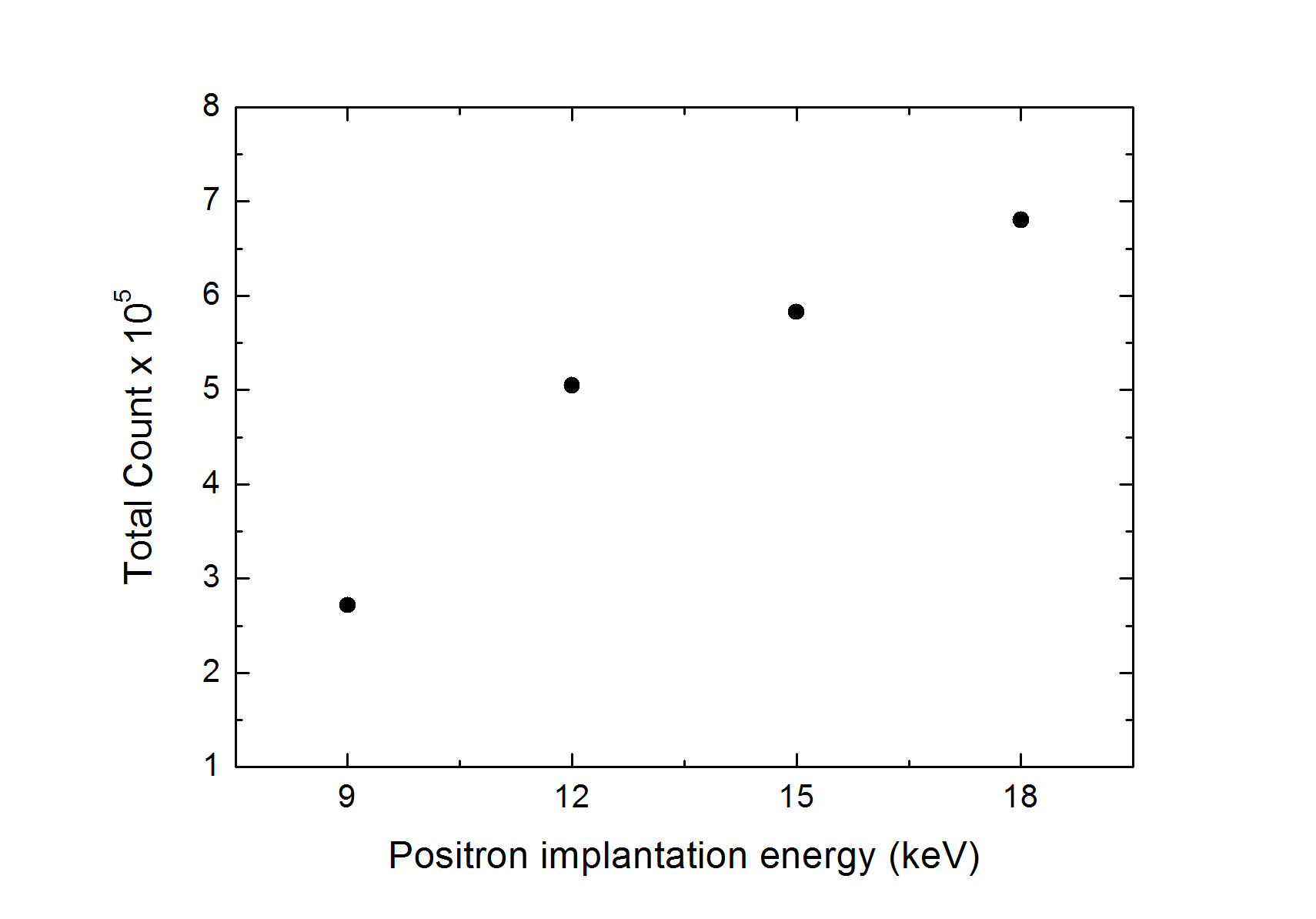}
\caption{Number of events measured by the emulsion detector as a function of the positron implantation energy. For the exposures at 3 and 6 keV no positrons were observed.}
\label{counts_vs_energy}
\end{center}
\end{figure}

\begin{figure}[h]
\begin{center}
\includegraphics[width=0.47\textwidth]{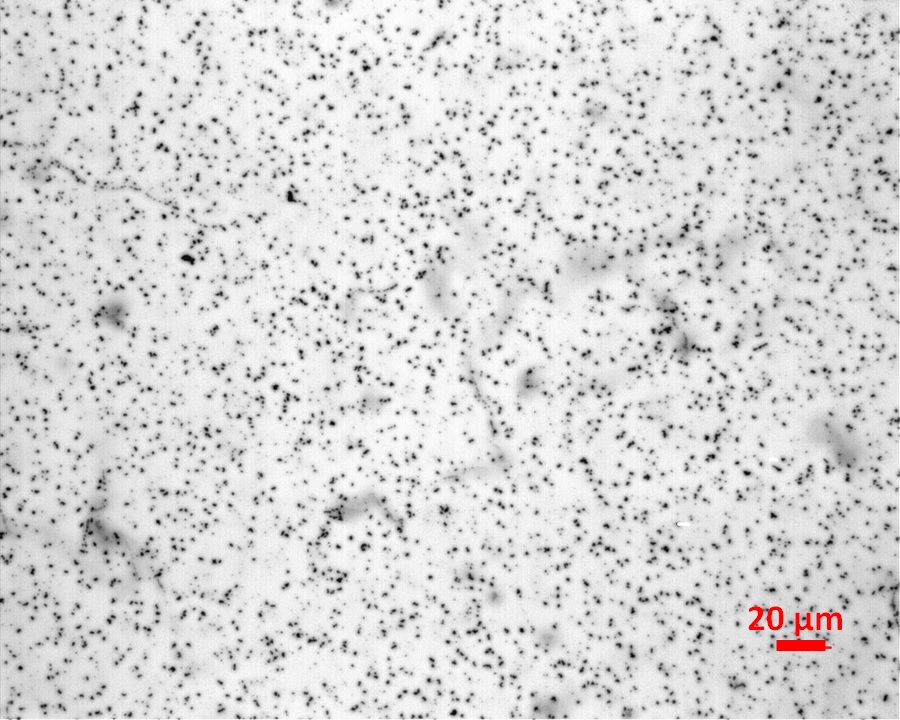}
\hspace{1mm}
\includegraphics[width=0.47\textwidth]{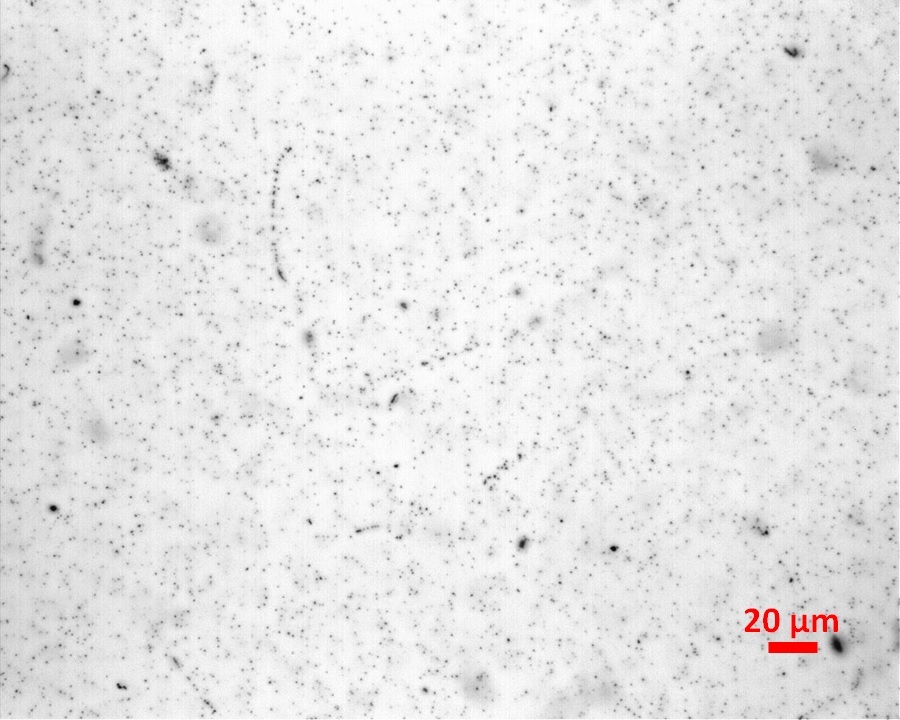}
\caption{Images of the center of the 15 keV positron spot (left) and of a region not exposed (right). The size difference between clusters created by positron interactions (signal) and thermal effects (background) is evident.}
\label{em_image}
\end{center}
\end{figure}

\indent
The number of positrons for each spot was evaluated under the assumptions of a gaussian distribution for the positron signal and of a linear background. The vertical beam profile histogram was then fitted and the resulting gaussian integrated to estimate the positron number. The total number of detected positrons as a function of the positron implantation energy is shown in Figure~\ref{counts_vs_energy}. The number of detected positrons after background subtraction ranges between $\sim$ 3-7$\times 10^5$ and the statistical error, evaluated on the integral of the gaussian fit of the signal, was estimated to be between 0.2\% - 1.2\%. No positrons were detected at energies of 3 and 6 keV. \\
\indent
As the background rate does not depend on the exposure time, this could be properly set to maximize the statistical significance of the signal. An improved background rejection could be obtained by considering also the size of reconstructed clusters, as it clearly appears in Figure~\ref{em_image}, where two images of the center of the 15 keV spot (left) and of a non exposed region (right) are shown. The positrons cross indeed very often more than one silver bromide crystal and after the emulsion development the resulting average cluster size is considerably larger than the size of the grains produced by thermal effects. This highly benefits interferometry experiments, where very low particle fluxes are expected. \\
\indent
The number of events increases with the implantation energy. In principle, this could be due to both the increased number of positrons reaching the emulsion layer, and to the enhancement of the detection efficiency with the energy. In order to compare the number of events detected by the emulsion and the effective number of positrons implanted into the emulsion for each positron energy, the positron implantation profile was estimated by means of an algorithm based on the formulation developed by Ghosh et al.~\cite{Ghosh} and Aers et al.~\cite{Aers1},~\cite{Aers2},~\cite{Aers3}. This semi-empirical method well reproduces the Monte Carlo stopping profiles in elemental multi-layers at the energy range of implanteted positrons investigated in the present work. The Ghosh-Aers formulation was adopted by virtue of its easiness and stability to small uncertainties of the material parameters. Moreover, it also accounts for the positron transmission and reflection in correspondence to each surface.
A fraction of positrons is either backscattered by the emulsion protection layer or annihilated in the same layer~\cite{Makinen} and the parameters used for the calculation depend on the properties of the multilayer materials, such as density and thickness. Even if their values for the protective and the emulsion layers had to be estimated from the literature~\cite{Dryzek}, the model was considered sufficiently realistic for our purposes. 
\begin{figure}[h]
\begin{center}
\includegraphics[width=0.48\textwidth]{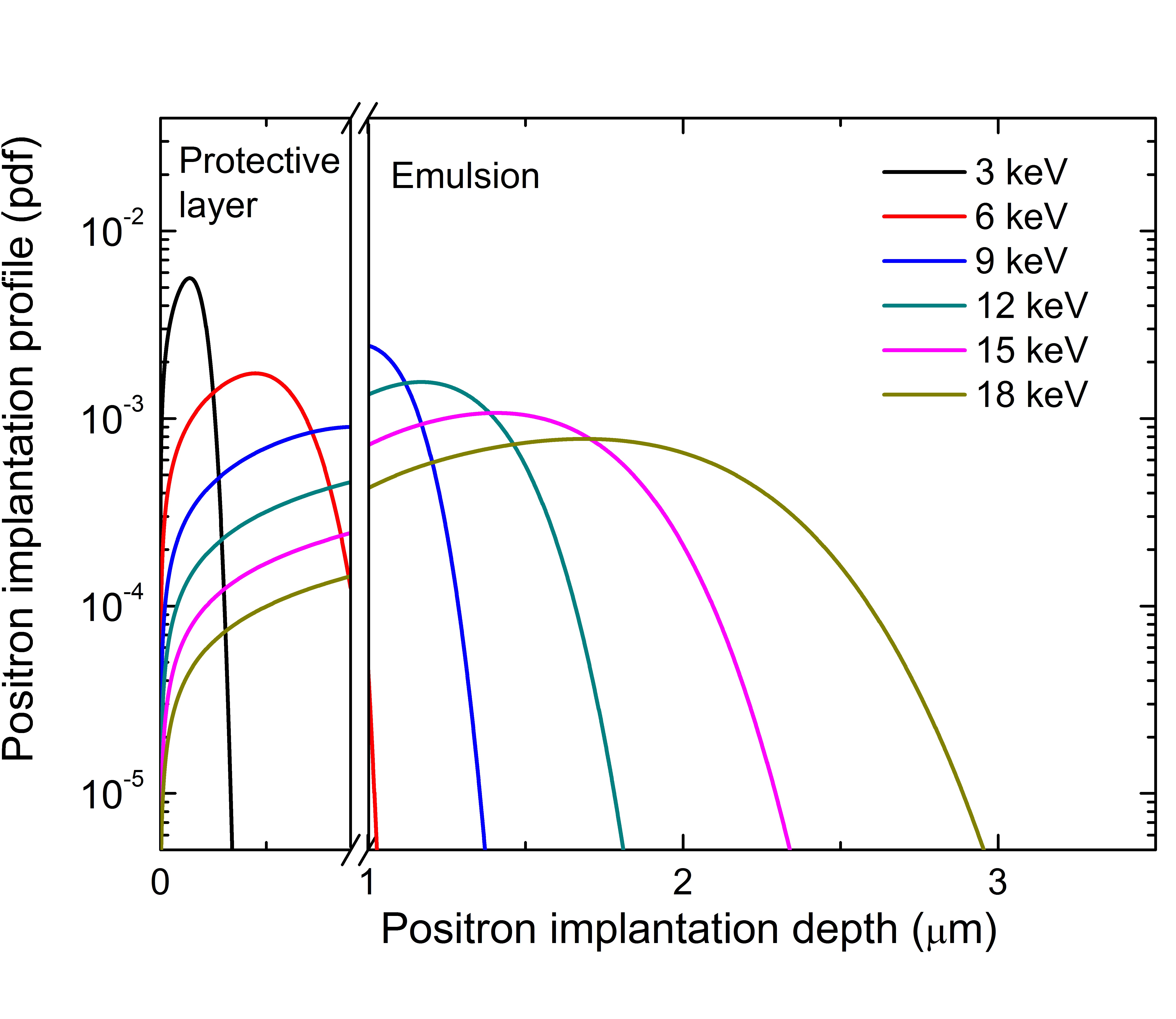}
\hspace{4mm}
\includegraphics[width=0.48\textwidth]{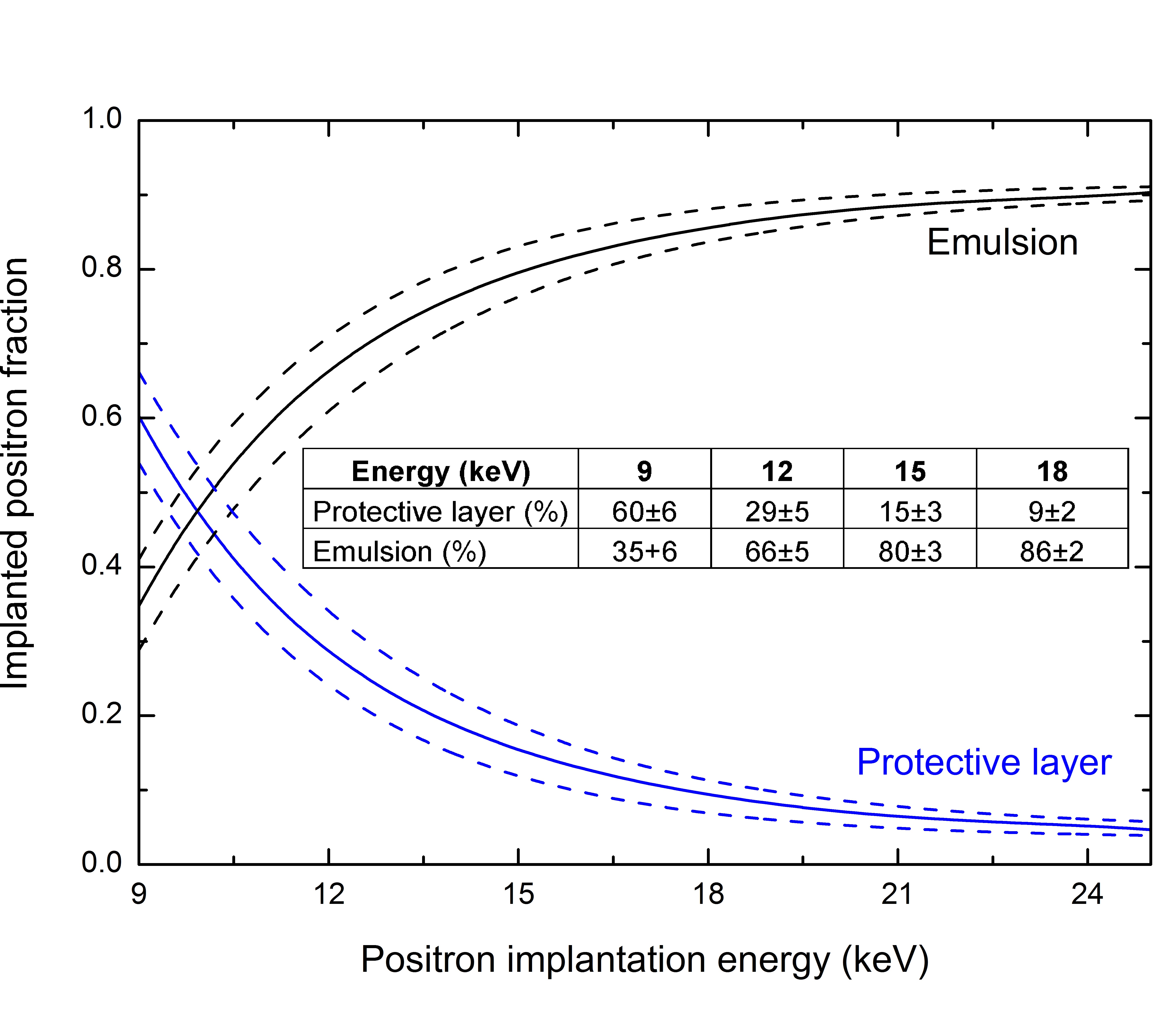}
\caption{Calculated positron implantation profiles (left) and fractions of positrons undergoing annihilation inside the protective and the emulsion layers as a function of the implantation energy (right).}
\label{implantation_profile}
\end{center}
\end{figure}

\begin{figure}[h]
\begin{center}
\includegraphics[width=0.7\textwidth]{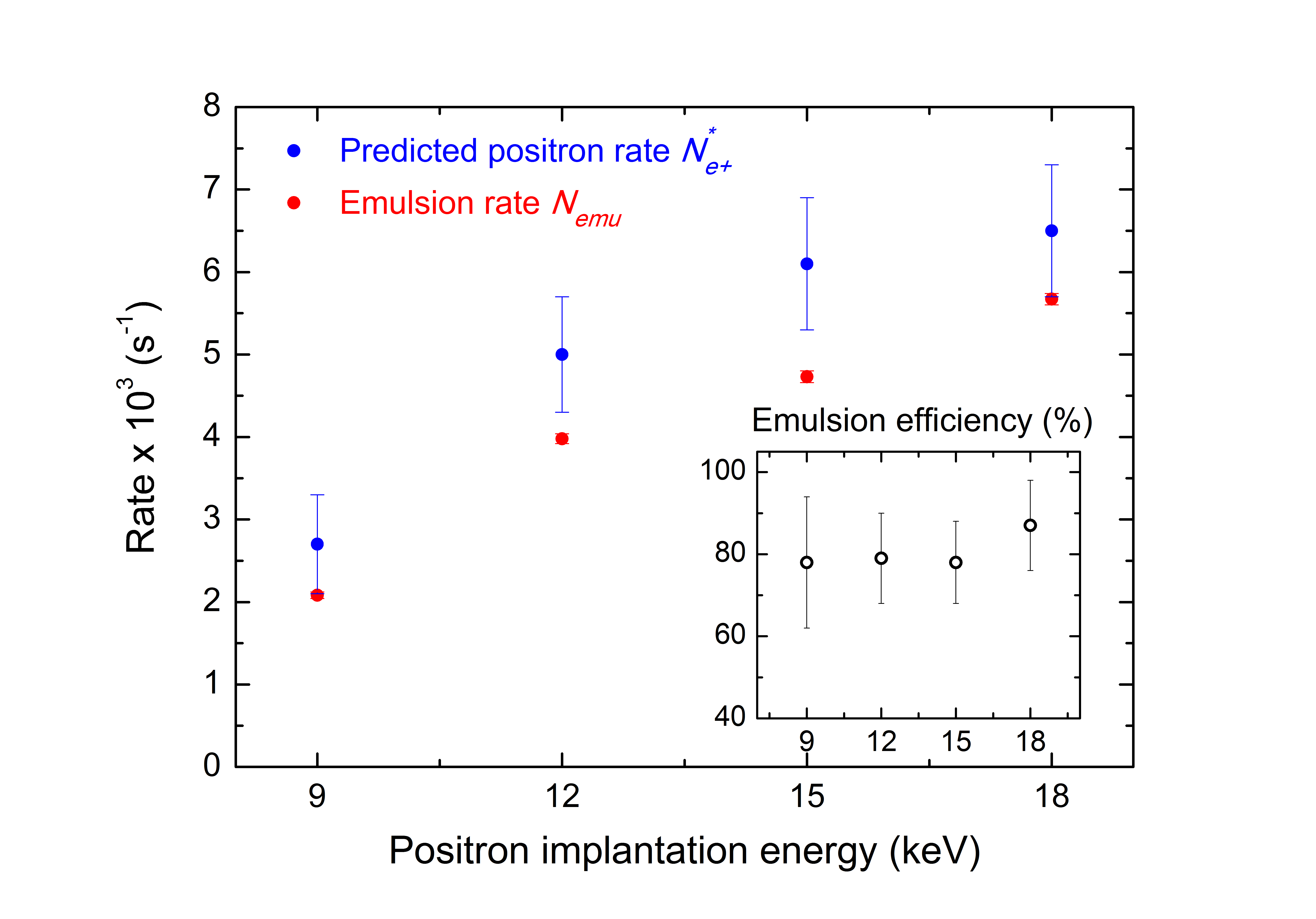}
\caption{A comparison between the predicted rate of implanted positrons and the rate detected by the emulsion detector as a function of the implantation energy. Emulsion efficiencies are shown in the embedded figure.}
\label{count_rate}
\end{center}
\end{figure}

\indent
Figure~\ref{implantation_profile} shows the calculated positron implantation profiles for the different energies considered in the present work (left) and the calculated fractions of positrons that annihilate inside the protective polymer layer and the emulsion layer $f_{emu}$ (right), corresponding to the area under the curves. Possible fluctuations of the emulsion protective layer thickness were also taken into account. As they could not be directly measured, a conservative 10$\%$ error was assumed reflecting in the uncertanties reported in the table. As already mentioned at the energy of 3 keV and 6 keV, positrons do not reach the emulsion layer and therefore the corresponding values are not listed in the table. The evaluated number of positrons backscattered at the surface of the protective polymer layer does not depend on the energy and was found to be $5\pm1\%$~\cite{Makinen}. The remaining 95$\%$ was considered for the calculation. \\
\indent
During emulsion exposure, the annihilation radiation (511 keV photons) from the implanted positrons was measured by means of two germanium detectors. The positron count below the annihilation peak was found to slightly decrease for increasing beam energy (the reduction with respect to the case of 9 keV being approximately $8\%$ at 18 keV). This variation is due to the presence of a tungsten collimator along the positron path (see  Figure~\ref{drawing}). Indeed at high implantation energies a fraction of positrons does not pass through the collimator hole. This was kept into account by properly reweighting the integrated counts on the emulsions at energies from 9 to 15 keV with the corresponding positron flux. Therefore, the values shown in Figure~\ref{counts_vs_energy} have been suitably rescaled. \\
\indent
The emulsion efficiency $\epsilon_{emu}$ for positron detection was calculated for each energy as the ratio between detected $N_{emu}$  and predicted $N^*_{e^+}$ positron rates. $N^*_{e^+}$ is calculated as $N^*_{e^+}$ = $f_{emu}N_{e^+}$ where $f_{emu}$ are the fractions of implanted positrons reported in Figure~\ref{implantation_profile} and $N_{e^+}$ is the positron flux of $(7.6\pm0.9)\times 10^{3}$ s$^{-1}$ estimated with the HPGe detectors at 18 keV. The obtained values for the efficiency are reported in Table~\ref{tb:eff}: within the errors it turns out to be independent from the positron implantation energy. A summary of the results is also shown in  Figure~\ref{count_rate}.
\begin{table}[h]
\begin{center}
 \begin{tabular}{|c|c|c|c|c|}
        \hline
  \bf Positron  & \bf Emulsion positron    & \bf Predicted positron   & \bf Emulsion rate   & \bf Efficiency   \\
  \bf Energy (keV) & \bf fraction $\mathbf{f_{emu}}$    & \bf rate  $\mathbf{N^*_{e^+} (10^3 s^{-1})}$& \bf $\mathbf{N_{emu} (10^3 s^{-1})}$   & $\mathbf{\epsilon_{emu}\times100\%}$   \\
 \hline
    \bf     9& 0.35$\pm$0.06 & 2.7$\pm$0.6 & 2.08$\pm$0.04 & 78$\pm$16  \\ \hline
    \bf   12& 0.66$\pm$0.06 & 5.0$\pm$0.7 & 3.98$\pm$0.06 & 79$\pm$11  \\ \hline
    \bf   15& 0.80$\pm$0.03 & 6.1$\pm$0.8 & 4.73$\pm$0.07 & 78$\pm$10  \\ \hline
    \bf   18& 0.86$\pm$0.02 & 6.5$\pm$0.8 & 5.67$\pm$0.07 & 87$\pm$11  \\ \hline
    \end{tabular}
\caption{Fraction of implanted positrons with corresponding predicted rate, measured rate by the emulsions and resulting efficiency for each value of implantation energy. }
\label{tb:eff}
\end{center}
\end{table}

\vspace{-6mm}
\section{Conclusions}
We studied the sensitivity of the emulsion detector using a mono-energetic positron beam at low energies, between 9 keV and 18 keV. For this purpose we developed and realised a new emulsion detector, using a gel with enhanced content of silver bromide crystals poured on a glass plate. Nuclear emulsions were found to be highly efficient at detecting positrons at the considered energy range and the efficiency was found to be independent from the positron implantation energy.  \\
\indent
The experimental methodology described in this paper paves the way to matter-wave interferometry with positrons. Future exposures with a better knowledge of the incident positron flux for an improved assessment of the emulsion detection efficiency are being considered together with the exposure of an emulsion detector with a thinner protective layer in order to detect positrons with lower energies. \\

\acknowledgments
The authors would like to acknowledge the contributions by the mechanical workshops at LHEP and LNESS. This work is supported  by the Swiss National Science Foundation Ambizione grant  PZ00P2$\_$154833.


\begin{thebibliography}{9}

\bibitem{emul_review2}
A. Ereditato, \emph{The Study of Neutrino Oscillations with Emulsion Detectors}, \emph{Adv. High Energy Phys.} \textbf{2013} (2013) 382172. 

\bibitem{AEgIS_emulsion1}
C. Amsler et al., \emph{ A new application of emulsions to measure the gravitational force on antihydrogen},
 \emph{JINST} \textbf{8} (2013) P02015. 

\bibitem{AEgIS_emulsion2}
S. Aghion et al., \emph{Prospects for measuring the gravitational free-fall of antihydrogen with emulsion detectors},
 \emph{JINST} \textbf{8} (2013) P08013. 

\bibitem{AEgIS_minimoire}
S. Aghion et al., \emph{A moir\'e deflectometer for antimatter}, \emph{Nat. Comm.} {\bf 5} (2014) 4538.

\bibitem{Sala_antimatter}
S. Sala et al., \emph{Matter-wave interferometry: towards antimatter interferometers}, 
\emph{J. Phys. B: At. Mol. Opt. Phys.} {\bf 48} (2015) 19002. 

\bibitem{ATmic}
A. Ariga and T. Ariga, \emph{Fast 4$\pi$ track reconstruction in nuclear emulsion detectors based on GPU technology},
 \emph{JINST} \textbf{9} (2014) P04002. 

\bibitem{Ghosh}
V. J. Ghosh, D. O. Welch and K. G. Lynn, \emph{Proc. Of the 5th Int. Workshop on Slow-Positron Beam Techniques for Solids and Surfaces}, Ed. E. H. Ottewitte (AIP, New York, 1993). 

\bibitem{Aers1}
G.C. Aers, P.A. Marshall, T.C. Leung, R.D. Goldberg, \emph{Defect profiling in multilayered systems  using mean depth scaling},
{\emph{Applied Surface Science} {\bf 85} (1995) 196}.

\bibitem{Aers2}
G.C. Aers, \emph{Simple scaling law for positron stopping in multilayered systems},
{\emph{Appl.\ Phys. \ Lett.} {\bf 64} (1994) 661}.

\bibitem{Aers3}
G.C. Aers, P.A. Marshall, T.C. Leung, R.D. Goldberg, \emph{Positron stopping profiles in multilayered systems},
{\emph{J.\ Appl.\ Phys.} {\bf 71} (1994) 1622}.

\bibitem{Makinen}
J. M\"akinen, S. Palko, J. Martikainen and P. Hautoj\"arvi, \emph{Positron backscattering probabilities from solid surfaces at 2-30 keV},
{\emph{ J. Phys.: Condens. Matter} {\bf 4} (1992) L503}.

\bibitem{Dryzek}
J. Dryzek, P. Horodek, \emph{Positron stopping profiles in multilayered systems},
{\emph{ Nucl.\ Inst.\ Met.} {\bf  B266} (2008) 4000}.


\end{thebibliography}
\end{document}